\documentclass[showpacs,preprintnumbers,amsmath,amssymb,twocolumn,superscriptaddress,prb]{revtex4}
\usepackage{times}
\usepackage{graphicx}
\usepackage{amsfonts}
\usepackage{amsmath, amsthm, amssymb}
\usepackage{dsfont}
\usepackage{subfigure}

\newcommand{\vk}{\mathbf{k}} 
\newcommand{\vp}{\mathbf{p}} 
\newcommand{\vpf}{\mathbf{p}_\text{F}}
\newcommand{\vecr}{\mathbf{r}}
\newcommand{\vq}{\mathbf{q}} 
\newcommand{\e}[1]{\mathrm{e}^{#1}}
\newcommand{\g}{\check{g}(\vpf,\vecr;\varepsilon,t)}
\newcommand{\G}{\check{G}(\vp,\vecr;\varepsilon,t)} 
\newcommand{\ug}{\underline{g}}
\newcommand{\uf}{\underline{f}}
\newcommand{\eg}{\textit{e.g.}}

\def\i{\mathrm{i}}

\begin{document}
\title[Quantum transport in a normal metal/odd-frequency superconductor junction]
{Quantum transport in a normal metal/odd-frequency superconductor junction}
\author{Jacob Linder}
\affiliation{Department of Physics, Norwegian University of
Science and Technology, N-7491 Trondheim, Norway}
\author{Takehito Yokoyama}
\affiliation{ Department of Applied Physics, Nagoya University, Nagoya 464-8603, Japan and CREST-JST}
\author{Yukio Tanaka}
\affiliation{ Department of Applied Physics, Nagoya University, Nagoya 464-8603, Japan and CREST-JST}
\author{Yasuhiro Asano}
\affiliation{Department of Applied Physics, Hokkaido University, Sapporo 060-8628, Japan}
\author{Asle Sudb{\o}}
\affiliation{Department of Physics, Norwegian University of
Science and Technology, N-7491 Trondheim, Norway}

\date{Received \today}
\begin{abstract}
Recent experimental results indicate the possible realization 
of a bulk odd-frequency superconducting state 
in the compounds 
CeCu$_2$Si$_2$, and CeRhIn$_5$. Motivated by this, we present a study of the quantum transport properties of a normal metal/odd-frequency superconductor junctions in a search for probes to unveil the 
odd-frequency symmetry. From the Eliashberg equations, we perform a quasiclassical approximation to account for the transport formalism of an odd-frequency superconductor with the Keldysh formalism. Specifically, we consider the tunneling charge conductance and tunneling thermal conductance. We find qualitatively distinct behaviour in the odd-frequency case as compared to the 
conventional even-frequency case, in both the electrical and thermal current. This 
serves as a useful tool to identify the possible existence of a bulk odd-frequency superconducting state.
\end{abstract}

\maketitle
\section{Introduction}
The symmetries of the superconducting order parameter with respect to orbital-, time-, and spin-space are 
governed by the Pauli principle. A wavefunction describing two electronic states must be totally antisymmetric 
under exchange of the particle-coordinates. This leads to a finite number of allowed combinations for the 
symmetries of the wavefunction. In a wide variety of superconductors ranging from those described 
with Bardeen-Cooper-Schrieffer/Eliashberg theory via spin-triplet superconductivity in $^3$He, to strong-coupling 
 superconductivity in high-$T_c$ cuprates, the wave function of Cooper-pairs is even in the frequency 
domain. For such even-frequency pairing, the wavefunction may be even or odd in space depending on whether the Cooper-pairs form spin-singlets or -triplets. 
However, more exotic types of pairings than what is found in this wide range of materials, are in principle permitted. 
\par
Recently, it was predicted that in a ferromagnet/superconductor structure, a so-called \textit{odd-frequency} 
pairing could take place \cite{bergeretPRL}. Thus, the Cooper pair wavefunction is symmetric under exchange 
of spatial- and spin-coordinates, but antisymmetric under exchange of time-coordinates. This state had been 
proposed to exist by Berezinskii \cite{berezinskii} a few decades earlier in the context of liquid $^3$He, 
and strong experimental evidence for odd-frequency pairing now exists \cite{Kaizer}. The 
study of such pairing in ferromagnet/conventional superconductor junctions has been addressed by a 
number of authors over the last years \cite{ferro}. Furthermore, it was very recently predicted  
\cite{tanakaPRLNEW,Eschrig,tanuma} that due to spatial variation of the pair potential near a normal/superconductor (N/S) 
junction, the odd-frequency pairing state can be induced even in a conventional ballistic N/S 
system without spin-triplet ordering. 
The generation
of different symmetry components and their effect on electrical transport in a 
normal/superconductor interface has also been studied in the diffusive 
limit \cite{Asano2007} in the context of the 
proximity effect in unconventional superconductors \cite{Proximity, nazarov_generalized}. 
\par
An issue that arises in the context of the odd-frequency pairing state, 
is if it can be realized in a 
{\it bulk} superconductor, i.e. without a proximity effect. 
There have been several theoretical proposals for this in strongly 
correlated systems up to now \cite{Balatsky,Coleman}. 
To explore an odd-frequency pairing state in heavy-fermion superconductors 
is an interesting topic, and an assessment of the experimental properties of CeCu$_2$Si$_2$, and CeRhIn$_5$ concluded that 
odd frequency pairing may be realized in these heavy-fermion compounds \cite{fuseya}. 
\begin{figure}[h!]
\centering
\resizebox{0.45\textwidth}{!}{
\includegraphics{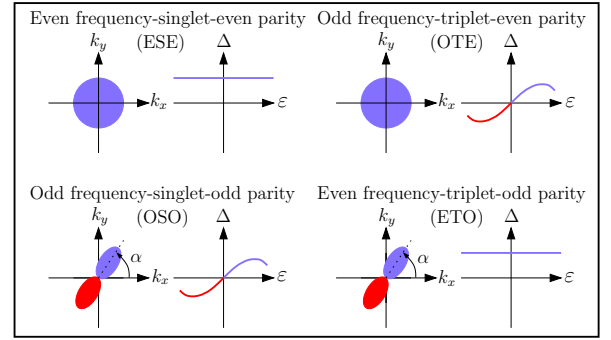}}
\caption{(Color online) Overview of the different symmetry states we will consider in the superconducting part 
of the clean, two-dimensional normal/superconductor junction.}
\label{fig:model}
\end{figure}

However, only 
a very limited amount of studies have addressed the issue of identifying the odd-frequency pairing 
state in a bulk superconductor so far \cite{Balatsky,Golubov2007, linder_yokoyamaPRL}. Hence, further clear-cut predictions 
are needed. 
\par
In this paper, we present the quantum transport properties of a normal metal/odd-frequency superconductor junction in 
the clean limit. We calculate the electrical and thermal conductances within the 
Blonder-Tinkham-Klapwijk (BTK) framework \cite{btk} taking account of the 
anisotropy of the pair potential \cite{tanaka}. Our starting point is the Eliashberg equations that take into account the frequency-dependence of the pair potential. This constitutes a wide 
range of experimental predictions, which are routinely used to characterize superconducting states 
\cite{characterize_1,characterize_2,characterize_3,characterize_4}. 
{\it Our main result is that the odd-frequency symmetry affects the charge (thermal) transport in an essential 
manner at low energies (temperatures)}. This provides a useful tool in identifying this highly unusual 
superconducting state.
\par
To elucidate the physics in a transparent manner, we employ a simple two-dimensional calculation 
in the clean limit. We approximate the superconducting gap with a step-function 
in space, which in the isotropic even-parity $s$-wave case should be an excellent approximation for 
low-transmission barriers. Since the low-transmission case probably is the most realistic scenario 
experimentally, we restrict our attention to this. In the anisotropic even-parity and 
odd-parity cases (corresponding e.g. to the high-T$_c$ superconductors and Sr$_2$RuO$_4$), the gap may 
undergo a severe depletion near the barrier even for low-transmission interfaces due to the formation 
of zero-energy states \cite{hu}. The method used in this paper may still be able to capture 
qualitative features of the transport properties even in those cases, just as in the case of the 
$d$-wave superconductors \cite{tanaka}. Our results are in fact consistent with recent 
findings \cite{tanakaPRLNEW} including a self-consistent solution of the 
spatial variation of the superconducting gap near the interface. 
\par
We will use 
boldface notation for 3-vectors, $\hat{\ldots}$ for $4\times4$ 
matrices, and $\underline{\ldots}$ for $2\times2$ matrices. Pauli-matrices in particle-hole$\times$spin (Nambu) space are denoted as $\hat{\rho}_i$, while Pauli-matrices in spin-space are written as $\underline{\tau}_i$. 
\section{Theoretical formulation}
\subsection{Equations for odd-frequency superconductivity}
The frequency-dependence of the superconducting order parameter may be naturally taken into account in the approach developed by Eliashberg \cite{eliashberg}, where details of the electron-boson interaction are taken seriously. This contrasts the usual weak-coupling picture where the pairing interaction is taken to be constant. For our purposes, the following Hamiltonian is an appropriate starting point:
\begin{align}
&\hat{H} = \sum_\alpha \int \text{d}\vecr \psi_\alpha(\vecr)^\dag H_f(\vecr) \psi_\alpha(\vecr) + \int \text{d}\vecr b^\dag(\vecr) H_b(\vecr)b(\vecr) \notag\\
&+\sum_\alpha\int\int \text{d}\vecr \text{d}\vecr' \mathcal{V}(\vecr-\vecr') \psi_\alpha^\dag(\vecr)\psi_\alpha(\vecr)[b(\vecr) + b^\dag(\vecr)],
\end{align}
where $H_f$ is the Hamiltonian for free fermions which we assume may be written as $H_f(\vecr) = -\frac{1}{2m}(\nabla -\i e\mathbf{A})^2 - \mu$, while $H_b$ is the Hamiltonian for free bosons.
Above, $\alpha$ denotes the spin index while $\psi$ and $b$ are fermion and boson operators, respectively.
Introducing the Fourier-transformation $b(\vecr) = \frac{1}{N} \sum_\vq b_\vq \e{-\i\vq\cdot\vecr}$, $B_\vq = b_\vq + b_{-\vq}^\dag,$
we obtain the Heisenberg equations of motion
\begin{align}
\i\partial_t \psi_\alpha(\vecr,t) &= H_f(\vecr)\psi_\alpha(\vecr,t) + \sum_\vq \zeta(\vecr,t,\vq) \psi_\alpha(\vecr),\notag\\
\i\partial_t \psi^\dag_\alpha(\vecr,t) &= -H_f^*(\vecr)\psi^\dag_\alpha(\vecr,t) - \sum_\vq \zeta(\vecr,t,\vq)\psi^\dag_\alpha(\vecr),
\end{align}
where $\zeta(\vecr,t,\vq) \equiv \mathcal{V}_\vq B_\vq(t) \e{-\i\vq\cdot\vecr}$ and $\mathcal{V}_\vq$ is the Fourier-transform of $\mathcal{V}(\vecr-\vecr')$. Note that $\mathcal{V}$ is \textit{not} the effective pairing potential between electrons. Having obtained the time-derivatives of the fermion operators, we may now calculate the equation of motion for the Green's functions. This procedure is standard and covered in \eg~Refs.~\onlinecite{serene, kopnin, rammer, zagoskin}. 
Taking into account the effect of the electron-boson interactions explicitly in the Hamiltonian naturally includes a frequency-dependence in the effective electron-electron interaction \cite{eliashberg} which is obtained by integrating out the bosonic degrees of freedom in the partition function. The effective electron-electron interaction mediated by a boson excitation may in general be written as
\begin{align}\label{eq:pair}
V(\vq,\Omega) &= \frac{2|\mathcal{V}_\vq|^2\omega_\vq}{\omega_\vq^2+\Omega^2},
\end{align}
where $\vq=\vk-\vk'$ and $\Omega=\omega-\omega'$ are the momentum and energy transfers, respectively, of the interaction process. 
Above, $\omega_\vq$ is the frequency of the boson propagator. Note that the pairing potential in Eq. (\ref{eq:pair}) is \textit{even} 
in $\Omega$, i.e. $V(\vq,\Omega) = V(\vq,-\Omega)$. The self-consistency equation for the order parameter quite generally has the 
structure 
\cite{Balatsky}
\begin{align}
\Delta(\vk,\omega) &\sim \sum_{\vk'\omega'} \frac{V(\vk-\vk',\omega-\omega')\Delta(\vk',\omega')}{\varepsilon_\vk'^2 + \omega'^2},
\end{align}
which may be re-written as 
\begin{align}
\Delta(\vk,-\omega) &\sim \sum_{\vk'\omega'} \frac{V(\vk-\vk',\omega-\omega')\Delta(\vk',-\omega')}{\varepsilon_\vk'^2 + \omega'^2},
\end{align}
by exploiting $V(\vq,\Omega) = V(\vq,-\Omega)$. The above equations show that both $\Delta(\vk,\omega) = \Delta(\vk,-\omega)$ and $\Delta(\vk,\omega) = -\Delta(\vk,-\omega)$ are possible solutions of the gap equation.
Therefore, although the pairing interaction is even in frequency, the gap $\Delta$ 
in principle may be both even or odd in frequency. In fact, it is in general a superposition of even- and odd-frequency components 
\cite{Balatsky,vojta}. Assuming that the energy transfer is small compared to the term containing the momenta in Eq. (\ref{eq:pair}), 
$|\Omega| \ll |\omega_\vq|$, one obtains a part of the pairing potential which is linear in $\omega$ and $\omega'$ and one that 
is quadratic in the same quantities \cite{Balatsky} . The former part is the necessary ingredient to obtain a superconducting 
order parameter that is odd-in-frequency. It is also possible to adopt a purely phenomenological approach to an odd-frequency 
superconductor by assuming the frequency-dependence of the gap {\it a priori} \cite{bunder}.
\par
Let us now consider the structure of the Green's function matrix for an odd-frequency superconductor. It is instructive to 
briefly mention the result for an ordinary BCS superconductor, which has an even frequency-singlet-even parity (ESE) 
symmetry. In the BCS case, one obtains
\begin{equation}\label{eq:bcs}
\Big(\i\frac{\partial}{\partial t_1} \hat{\rho}_3 - \hat{\xi} - \hat{\Delta}(\vecr_1)\Big)\hat{G}^\text{R}(1,2) = \delta(1-2)\check{1}.
\end{equation}
Assuming a homogeneous and isotropic system where the Green's function only depends on the relative coordinates $t=t_1-t_2$ and $\vecr = \vecr_1-\vecr_2$, and where $\hat{\Delta}(\vecr_1) = \hat{\Delta}$ is a constant, one may Fourier-transform Eq. (\ref{eq:bcs}) according to $\hat{G}^\text{R}(\vp,\varepsilon) = \int\int\text{d}\vecr\e{-\i\vp\vecr}\text{d}t\e{\i \varepsilon t} \hat{G}^\text{R}(\vecr,t),$
where $\varepsilon$ and $\vp$ is the quasiparticle energy measured from Fermi level and momentum, respectively. We then obtain 
\begin{equation}
(\varepsilon\hat{\rho}_3 - \hat{\xi}_\vp - \hat{\Delta})\hat{G}^\text{R}(\vp,\varepsilon) = \hat{1},
\end{equation}
which upon matrix inversion yields the well-known BCS solution. The quasiclassical Green's 
functions $\g$ is obtained from the Gor'kov Green's functions $\G$ by integrating out the dependence on kinetic 
energy, assuming that $\check{G}$ is strongly peaked at Fermi level,
\begin{equation}\label{eq:quasiclassical}
\g = \frac{\i}{\pi} \int \text{d}\xi_\vp \G.
\end{equation}
The above assumption is typically applicable to superconducting systems where the characteristic length scale of the 
perturbations present, namely superconducting coherence length, is much larger than the Fermi wavelength. The 
corresponding characteristic energies of such phenomena must be much smaller than the Fermi energy $\varepsilon_\text{F}$. 
The quasiclassical Green's functions may be divided into an advanced (A), retarded (R), and Keldysh (K) component, each 
of which has a $4\times4$ matrix structure in the combined particle-hole and spin space.  One has that
\begin{equation}
\check{g} = \begin{pmatrix}
\hat{g}^\text{R} & \hat{g}^\text{K}\\
0 & \hat{g}^\text{A} \\
\end{pmatrix},
\end{equation}
where the elements of $\g$ read
\begin{equation}
\hat{g}^\text{R,A} = \begin{pmatrix}
\ug^\text{R,A} & \uf^\text{R,A}\\
-\tilde{\uf}^\text{R,A} & -\tilde{\ug}^\text{R,A} \\
\end{pmatrix},\;
\hat{g}^\text{K} = \begin{pmatrix}
\ug^\text{K} & \uf^\text{K}\\
\tilde{\uf}^\text{K} & \tilde{\ug}^\text{K} \\
\end{pmatrix}.
\end{equation}
The quantities $\ug$ and $\uf$ are $2\times2$ spin matrices, with the structure
\begin{equation}
\ug = \begin{pmatrix}
g_{\uparrow\uparrow} & g_{\uparrow\downarrow} \\
g_{\downarrow\uparrow} & g_{\downarrow\downarrow} \\
\end{pmatrix}.
\end{equation}
Due to internal symmetry relations between these Green's functions, all of these quantities are not independent. In particular, the 
tilde-operation is defined as
\begin{equation}
\tilde{f}(\vpf,\vecr;\varepsilon,t) = f(-\vpf,\vecr;-\varepsilon,t)^*.
\end{equation}
For a bulk $s$-wave superconductor, the retarded part may be expressed in terms of the normal and anomalous Green's functions $g$ and $f$ as follows:
\begin{align}\label{eq:bcsbulk}
\hat{g}^\text{R} = \begin{pmatrix}
g\underline{1} & f\i\underline{\tau_2}\e{\i\chi}\\
f\i\underline{\tau_2}\e{-\i\chi} & -g\underline{1}\\
\end{pmatrix},\;
\end{align}
\par
Here, $\chi$ is the globally broken U(1) phase associated with the spontaneous symmetry breaking of the superconducting state. In 
the odd-frequency case, however, one finally arrives at
\begin{equation}\label{eq:final}
[\varepsilon\hat{\rho}_3 - \hat{\xi}_\vp - \hat{\Delta}(\varepsilon)]\hat{G}^\text{R}(\vp,\varepsilon) = \hat{1},
\end{equation}
where now $\hat{\Delta}(\varepsilon)$ is the odd-frequency gap matrix. Note that Eq. (\ref{eq:final}) is equivalent to the well-known Eliashberg equation. The structure of the Green's function for an odd-frequency superconductor may be different from Eq. (\ref{eq:bcsbulk}) depending on the spin-symmetry. For instance, the bulk Green's function matrix for an odd-frequency spin-triplet even-parity superconductor has the structure:
\begin{align}
\hat{g}^\text{R} = \begin{pmatrix}
g\underline{1} & f\underline{\tau_1}\e{\i\chi}\\
-f\underline{\tau_1}\e{-\i\chi} & -g\underline{1}\\
\end{pmatrix},\;
\end{align}
Performing a quasiclassical approximation on Eq. (\ref{eq:final}) yields the Eilenberger equation, which reduces to the Usadel equation in the dirty limit. Note that for both even- and odd-frequency superconducting order parameters, the \textit{pairing interaction itself} is always even in the frequency coordinate. 
\par
A quite general formalism for treating quantum transport in non-uniform superconducting systems, \textit{e.g.}  normal/superconductor heterostructures, has been developed by Tanaka and co-workers \cite{tanaka_develop}. For instance, the conductance spectra of a normal/superconductor junction may be obtained along the lines of Refs.~\onlinecite{tanaka_develop, yokoyama_thermal} by numerically solving the Usadel equation using Nazarov's generalized boundary conditions \cite{nazarov_generalized}. Interestingly, taking the limit $R_d\to 0$ and $\theta\to 0$ in this formalism, where $R_d$ represents the resistance of the normal metal region and $\theta$ is a measure of the proximity effect, leads to the well-known expression for the conductance obtained in the BTK-formalism \cite{btk}. This may be seen specifically for the electrical conductance by consulting Eqs. (15) and  (16) in Ref.~\onlinecite{tanaka_develop}, and for the thermal conductance in Eq. (19) of Ref. ~\onlinecite{yokoyama_thermal} Therefore, since the above treatment of the Eliashberg equation shows that the odd-frequency dependence of the gap may be taken into account simply by substituting $\Delta\to\Delta(\varepsilon)$, quantum transport for an odd-frequency superconductor can be treated in the BTK-formalism by performing the same substitution. However, the derivation of the Bogolioubov-de Gennes equation for odd-frequency superconductivity is challenging since it is not obvious how to take into account the strong retardation effects of the pairing potential.   

\subsection{Transport formalism}
We adopt the Keldysh formalism using Nazarov's generalized boundary conditions \cite{nazarov_generalized} to obtain the electrical and thermal conductance for odd-frequency superconductors. We assume, without loss of generality, that the gap $\Delta(\varepsilon,\vartheta)$ has an opposite-spin pairing symmetry in both the singlet and triplet case. To encompass accessible experimental techniques, we will focus on two experimentally accessible quantities 
that encode how the odd-frequency pairing symmetry is manifested in transport properties: namely, the normalized 
charge-conductance $G(eV)$ for $T=0$ and the thermal-conductance $\kappa(T)$. The procedure for obtaining these quantities is treated in detail in Refs.~\onlinecite{tanaka_develop, yokoyama_thermal, yokoyama_magn}. In the limit of zero resistance in the normal part and vanishing proximity effect, one finds
\begin{align}
G &= \frac{1}{G_\mathrm{N}} \int^{\pi/2}_{-\pi/2} \text{d}\vartheta \cos\vartheta \Gamma_+(eV,\vartheta),\notag\\
\kappa &= \int^{\pi/2}_{-\pi/2} \int^\infty_{-\infty} \text{d}\vartheta\text{d}\varepsilon \frac{\varepsilon^2\beta^2\Gamma_-(eV,\vartheta)}{4\Delta_0\cosh^2(\beta\varepsilon)(\cos\vartheta)^{-1}},\notag\\
\end{align}
where $G_\text{N}$ is the normal-state conductance and we have defined
\begin{align}
&\Gamma_\alpha(\varepsilon,\vartheta) = 1 + \alpha\Bigg| \frac{4\Omega_-\tilde{\Omega}_+\e{-\i\gamma_+}}{\Omega_+\Omega_-(4-Z_\vartheta^2) + Z_\vartheta^2\tilde{\Omega}_+\tilde{\Omega}_-\e{\i(\gamma_--\gamma_+)}} \Bigg|^2 \notag\\
&- \Bigg|\frac{2[\Omega_+\Omega_-(2+Z_\vartheta) - Z_\vartheta \tilde{\Omega}_+\tilde{\Omega}_-\e{\i(\gamma_--\gamma_+)}]}{\Omega_+\Omega_-(4-Z_\vartheta^2) + Z_\vartheta^2\tilde{\Omega}_+\tilde{\Omega}_-\e{\i(\gamma_--\gamma_+)}}-1\Bigg|^2,
\end{align}
Above, we have introduced $\vartheta_+ = \vartheta,$ $\vartheta_- = \pi-\vartheta$, and $\Omega_\pm = \sqrt{(1 + \text{sign}(\varepsilon)/g_\pm)/2}$,
where $\text{sign}(\varepsilon)\to-\text{sign}(\varepsilon)$ for $\Omega\to \tilde{\Omega}$. 
The phase of the superconducting gap is contained in the factor $\e{\i\gamma_\pm} = \e{\i\gamma(\vartheta_\pm)} = f_\pm/|f_\pm|$. 
The quantities $g_\pm$ and $f_\pm$ are the asymptotic values of the normal and anomalous Green's functions of the odd-frequency 
superconductor in a gauge where the superconducting order parameter is real: $g_\pm = \varepsilon/\sqrt{\varepsilon^2 - |\Delta(\varepsilon,\vartheta_\pm)|^2},$ $f_\pm = \Delta(\varepsilon,\vartheta_\pm)/ \sqrt{ |\Delta(\varepsilon,\vartheta_\pm)|^2- \varepsilon^2 }$.
We have introduced $Z_\vartheta = -\i Z/\cos\vartheta$, where $Z$ denotes the strength of the scattering potential near the barrier. In what follows, we fix 
$Z=3$, corresponding to a typical low-transparency barrier which is experimentally realistic. Note that in the expression for $\kappa$, we have considered the linear response regime for a small temperature-gradient in the system and introduced $\beta=1/T$ where $T$ is the temperature of the reservoirs.
\begin{widetext}
\begin{figure}[h!]
\centering
\resizebox{0.95\textwidth}{!}{
\includegraphics{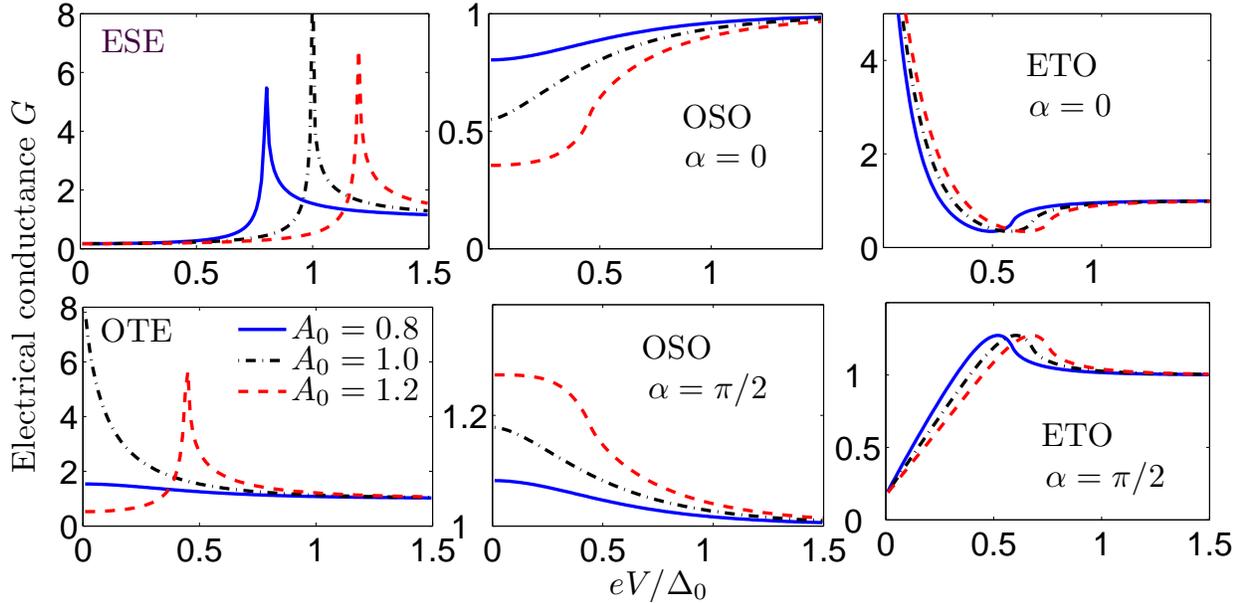}}
\caption{(Color online) Plot of $G$ for isotropic, even parity superconductors and odd 
parity superconductors for both even and odd frequency pairing. }
\label{fig:charge}
\end{figure}
\end{widetext}
\section{Results and discussion}
Depending on the symmetries with respect to sign 
inversion of frequency and momentum, 
corresponding to $\varepsilon\to(-\varepsilon)$ and $\vartheta\to\pi+\vartheta$, the gap may be classified 
as seen in Tab. \ref{tab:sym}. In each case, we will model the gap $\Delta(\varepsilon,\vartheta,T)$ as 
illustrated in the table. In the angular dependence of the odd-parity gaps, $\alpha$ denotes the 
misorientation angle between the antinodes and the interface normal (see Fig. \ref{fig:model}). The 
motivation for modelling the frequency-dependence of the superconducting gap as we have done in 
Tab. \ref{tab:sym} is that it features the low-energy behaviour of the proximity-induced odd-frequency 
gap in dirty ferromagnet/superconductor structures \cite{fominov2} and that it exhibits a similar energy 
dependence to the gap seen in strongly correlated electron systems considered in Ref.~\onlinecite{fuseya}. 
\begin{table}
\caption{Overview of the specific gap forms we will consider in this paper. We 
model the temperature-dependence of $A(T)$ with $A(T) = A_0\tanh(1.74\sqrt{T_c/T - 1})$, and 
$T_c = \Delta_0/1.76$.}
	\vspace{0.15in}
	\begin{tabular}{ccc}
	  	 \hline
	  	 \hline
		   Symmetry & Specific gap form $\Delta(\varepsilon,\vartheta,T)$\\
	  	 \hline
	  	 ESE &   $A(T)\Delta_0$\\
	  	 OTE &   $A(T)\varepsilon/[1 + (\varepsilon/\Delta_0)^2] $\\	 
	  	 OSO &  $A(T)\varepsilon\cos(\vartheta-\alpha)/[1 + (\varepsilon/\Delta_0)^2]$\\
	  	 ETO &  $A(T)\Delta_0\cos(\vartheta-\alpha)/[1 + (\varepsilon/\Delta_0)^2]$\\	  
	  	 \hline
	  	 \hline
	\end{tabular}
	\label{tab:sym}
\end{table}
\par
Recently, it was demonstrated that the odd-frequency pairing is quite generally induced near the 
normal/superconductor interface by a fully self-consistent calculation of the superconducting correlations 
\cite{tanakaPRLNEW}. In an ETO superconductor with $\alpha=0$, corresponding to perfect formation of 
zero-energy states, an OTE pairing is induced near the surface. Thus, the formation of zero-energy states 
may  be re-interpreted as a manifestation of the odd-frequency superconductivity near the interface. The 
odd-frequency symmetry may permit the existence of gapless single-particle 
excitations at Fermi level. On the other hand, when the nodal direction is parallel to the interface 
normal ($\alpha=\pi/2$), only the even-frequency states exist at the interface.
\par
In a similar manner, the OSO pairing state can be induced near the interface of a clean normal/superconductor junction 
when the superconductor has an ESE symmetry. One may also apply this discussion to bulk odd-frequency superconducting 
states. In this scenario, the ETO (ESE) pairing can be induced at the interface for an OTE (OSO) bulk superconductor 
\cite{tanakaPRLNEW}. This should have clearly observable consequences for the quantum transport properties of a 
normal/odd-frequency superconductor junction. We now proceed to investigate this in further detail.
\par
\begin{widetext}
\begin{figure}[h!]
\centering
\resizebox{0.95\textwidth}{!}{
\includegraphics{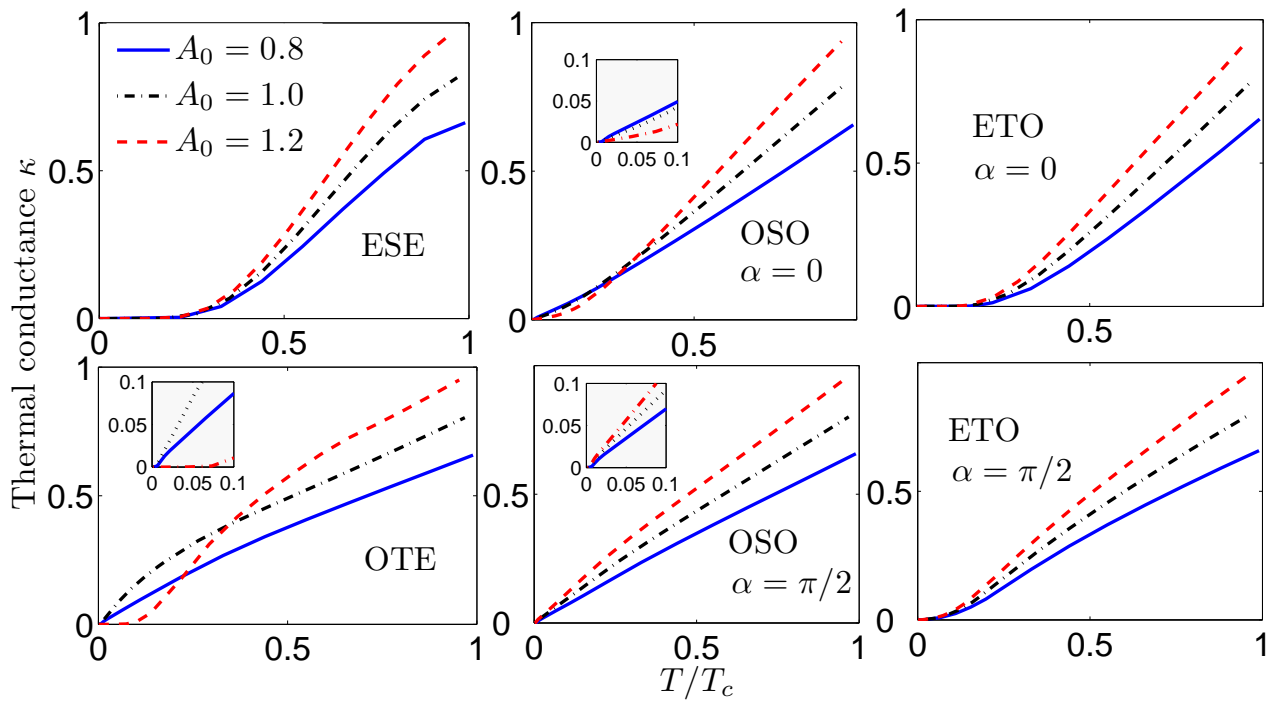}}
\caption{(Color online) Plot of $\kappa$ for isotropic, even parity superconductors and odd parity superconductors for both even and odd frequency pairing. A power-law dependence with exponent $\simeq1$ 
is observed for both of the odd frequency symmetries (see insets).}
\label{fig:heat}
\end{figure}
\end{widetext}
Consider first the left column of Fig. \ref{fig:charge} where we have plotted $G$ as a function of 
bias voltage for the even-parity symmetries. In the even-frequency case, the usual singularity at 
$eV=A_0\Delta_0$ is present. In the odd-frequency case, we see a qualitatively different behaviour 
of the conductance. First of all, $G$ at low bias voltage is greatly enhanced compared to the 
even-frequency case for $A_0<1$, and the formation of a zero-bias conductance peak is clearly seen at 
$A_0=1$. For $A_0>1$, the conductance is similar to the even-frequency case for a reduced value of the 
gap. This may be understood as follows. For $A_0 < 1$, the inequality $\varepsilon>\Delta(\varepsilon)$ 
is satisfied for all $\varepsilon$ with our choice of gaps (Tab. \ref{tab:sym}). This corresponds to gapless superconductivity. For $A_0>1$, the gap becomes 
larger than $\varepsilon$ below a certain (finite) value of $\varepsilon$, similar to the BCS-gap. 
\par
The middle and right columns of Fig. \ref{fig:charge} shows $G$ as a function of bias voltage for the odd-parity 
symmetries. In the OSO case, a gap like structure is seen at $\alpha=0$. This is because ESE pairing is 
induced near the interface due to the sign change of the pair potential. \cite{tanakaPRLNEW} This ESE 
pairing is responsible for the gap like structure of the conductance spectra, similar to the ESE case in 
Fig. 2. In contrast, OSO pairing remains  near the interface at $\alpha=\pi/2$. Thus, a zero-bias 
conductance peak is seen. On the other hand, in the ETO case at $\alpha=0$, a zero-bias conductance peak 
appears due to the induced OTE pairing near the interface\cite{tanakaPRLNEW}, similar to the OTE case in 
Fig. 2. At $\alpha=\pi/2$, ETO pairing survives near the interface  and hence the even frequency character of 
the pair amplitude results in a V-like shape of the spectra. Interestingly, OSO and ETO cases have the 
opposite tendency although their $\vartheta$-dependencies are the same. Furthermore, the sign change of the 
gap produces a qualitative difference in  the spectra between OTE and OSO with $\alpha=0$ junctions.  
Thus, $G$ is phase sensitive not only in even frequency superconductor junctions \cite{tanaka}, but also 
in odd frequency superconductor junctions. 
\par
We next investigate the thermal conductance $\kappa$, shown in Fig. \ref{fig:heat}. The left column corresponds 
to the even parity case, where the usual exponential dependence on $T$ is recovered for the ESE case \cite{andreev}. In the 
OTE case, $\kappa$ mimics the ESE case for $A_0>1$ just as for the charge conductance. Otherwise, power-law 
dependence with exponent $\simeq1$ is observed due to the node of the gap at zero energy. Thus, the nodes 
in the frequency domain of an isotropic odd-frequency superconductor causes $\kappa$ to behave like in an 
anisotropic even-frequency superconductor. In the middle and right columns of Fig. \ref{fig:heat}, we give $\kappa$ in 
the odd parity case. The well-known result of exponential dependence for $\alpha=0$ is recovered in the 
ETO case. The OSO case again displays power-law behaviour similar to the OTE case for $A_0<1$. However, 
the exponential dependence again occurs for $A_0>1$ in the OSO case with $\alpha=0$. When $\alpha=\pi/2$, 
there is exclusively power-law dependence, with exponent $\simeq1$. While the OTE case only has nodes in 
energy, the OSO case has both nodes in energy and in $\vk$-space, but this does not appear to influence 
the exponent of the power-law dependence.

\section{Summary}
In summary, we have studied quantum transport in a normal metal/superconductor junction, considering how a 
bulk odd frequency symmetry in the superconductor is manifested in the electrical and thermal conductance of 
the junction. The odd frequency symmetry is found to display qualitatively 
distinct behaviour from the even frequency case. This reflects the fact that the electrical conductance 
is sensitive to the presence of odd frequency pairing at the interface, whereas the low temperature behavior 
of the thermal conductance reflects the node of the gap in the frequency domain. Moreover, one may distinguish 
the even and odd parity cases for an odd frequency symmetry (OTE and OSO, respectively) by means of their 
different characteristic tunneling spectra. Our predictions should be useful for a wide range of experimental 
techniques, and are thus a helpful tool in identifying the possible existence of bulk odd frequency superconductors, 
with 
CeCu$_2$Si$_2$, and CeRhIn$_5$ currently presenting themselves as the most promising 
candidates.

\acknowledgments
J.L. and A.S. were supported by the Research Council of Norway, Grants No. 158518/431 
and No. 158547/431 (NANOMAT), and Grant No. 167498/V30 (STORFORSK). T.Y. acknowledges support by JSPS.  T.Y and 
Y.T were supported by Grant-in-Aid for Scientific Research (Grant Nos. 17071007) from the Ministry of Education,
Culture, Sports, Science and Technology of Japan. The authors acknowledge A. Balatsky for helpful comments. Y.T. would like to thank K. Miyake, H. Kohno and Y. Fuseya
for their valuable dicussions. J.L. acknowledges K. Yada for clarifying comments on the odd-frequency pairing 
potential.

\end{document}